\documentclass[aps,pre,twocolumn,groupedaddress]{revtex4-1}

\usepackage[english]{babel}

\usepackage[utf8]{inputenc}
\usepackage{times,mathptmx}
\usepackage{amsmath,amssymb}

\usepackage{graphicx} %

\usepackage{url}
\usepackage{xcolor}

\IfFileExists{./createPlots.tex}
{%
	\input{colors}
	\input{tikzSetup}
}{}

\newcommand{\includetikz}[2][]{\IfFileExists{./createPlots.tex}{\input{plots/#2}}{\includegraphics[#1]{#2}}}

\graphicspath{
	{./cache/}
}

\usepackage{bm}

\usepackage{xcolor}
\usepackage{prettyref}
\newrefformat{fig}{Fig.\,\ref{#1}}
\newrefformat{sec}{sect.\,\ref{#1}}
\usepackage{siunitx}

\newcommand{\re}{\mathsf{Re}}

\mathchardef\mhyphen="2D
\begin{document}
\title{Particle pairs and trains in inertial microfluidics}

\author{Christian Schaaf}
\email{christian.schaaf@tu-berlin.de}
\author{Holger Stark}
\affiliation
{Institut f\"ur Theoretische Physik, Technische Universit\"at Berlin, Hardenbergstr. 36, 10623 Berlin, Germany}
\date{\today}
\begin{abstract}
Staggered and linear multi-particle trains constitute characteristic structures in inertial microfluidics. 
Using lattice-Boltzmann simulations, we investigate their properties and stability, when flowing through microfluidic channels. 
We confirm the stability of cross-streamline pairs by showing how they contract or expand to their equilibrium axial distance. 
In contrast, same-streamline pairs quickly expand to a characteristic separation but even at long times slowly drift apart.
We reproduce the distribution of particle distances with its characteristic peak as measured in experiments.

Staggered multi-particle trains initialized with an axial particle spacing larger than the equilibrium distance contract non-uniformly due to collective drag reduction. 
Linear particle trains, similar to pairs, rapidly expand toward a value about twice the equilibrium distance of staggered trains and then very slowly drift apart non-uniformly. 
Again, we reproduce the statistics of particle distances and the characteristic peak observed in experiments. 
Finally, we thoroughly analyze the damped displacement pulse traveling as a microfluidic phonon through a staggered train and show how a defect strongly damps its propagation.
\end{abstract}
\maketitle

\section{Introduction}

Since the discovery of inertial focusing by Segr\'e and Silberberg~\cite{segre_behaviour_1962a}, inertial microfluidics has evolved into a mature research field with immense potential for biomedical applications \cite{zhang_fundamentals_2015,amini_inertial_2014,stoecklein_nonlinear_2019}.
At higher 
densities particles do not only move to an equilibrium position in the channel cross section but also form 
regular trains along the channel axis~\cite{lee_dynamic_2010,kahkeshani_preferred_2016,bazaz_computational_2020}.
This feature of inertial microfluidics is particular useful for counting~\cite{bhagat_inertial_2010,deng_inertial_2017,tang_microfluidic_2017}, sorting~\cite{bhagat_continuous_2008,mach_continuous_2010,li_size-based_2018,haddadi_separation_2018}, 
or manipulating cells~\cite{edd_controlled_2008,dhar_functional_2018}.
While 
trains of particles were already observed by Segr\'e and Silberberg~\cite{segre_behaviour_1962a}, the first  systematic analysis was done by Matas
\textit{et al.} about 40 years later~\cite{matas_trains_2004}.
This triggered further research on particle trains, which we describe below, in order to understand their occurrence more thoroughly.
In this article we contribute with our simulation study to the understanding of staggered and linear multi-particle trains including particle pairs (see \prettyref{fig:rigid_multi_setup}).

\subsection{Staggered and linear multi-particle trains}

\begin{figure}%
\begin{center}
	\includetikz{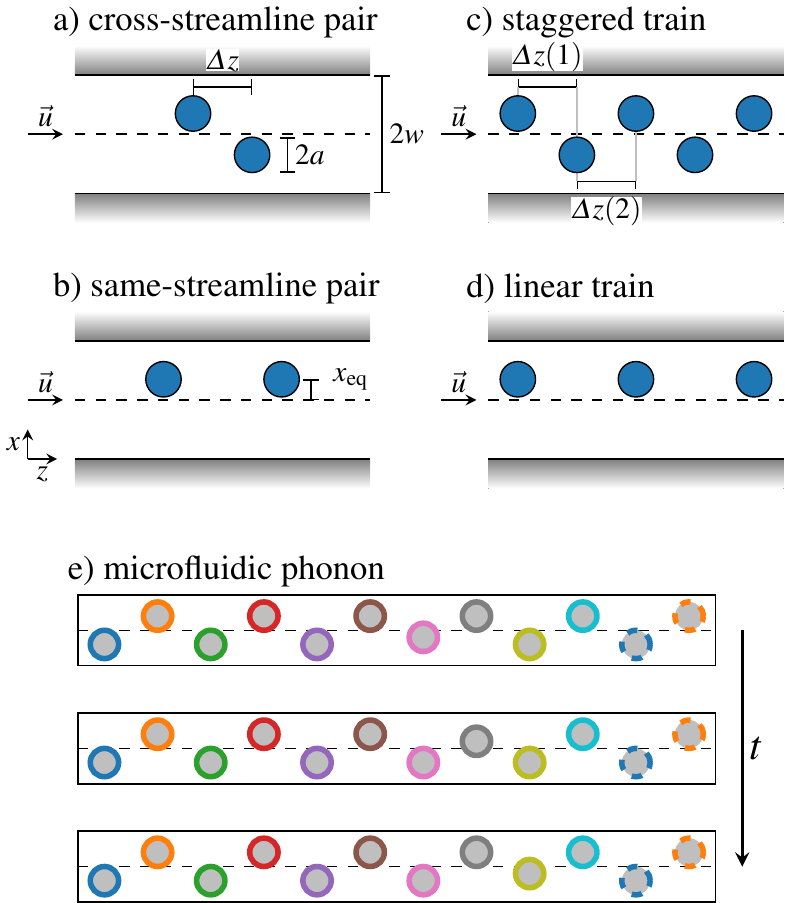}
\end{center}
	\caption{
		Illustrations of the multi-particle structures discussed in this work.
		Two	particles either form cross-streamline\,(a) or same-streamline\,(b) 
	        pairs.
	    Several particles form the corresponding
	    staggered 
	    \,(c) or linear trains\,(d).
        $x_{\mathrm{eq}}$ gives the equilibrium distance from the channel centerline, $\Delta z$ the distance between two neighboring particles, and the arrows with $\vec{u}$ indicate the flow direction.
        (e) A microfluidic phonon in a staggered particle train is triggered by moving one 
       particle toward the channel center (pink particle in the upper channel). 
       It
        moves faster than the rest of the train and 
        thereby a displacement pulse travels through the train.
        In the middle and bottom channel the maximal displacements of the gray and yellow particles are shown. The arrow indicates increasing time.
	}
	\label{fig:rigid_multi_setup}
\end{figure}

The first experiments used cylindrical tubes, where 
particles focus onto an annulus.
In Ref.\ \cite{matas_trains_2004} Matas \textit{et al.} studied particles with a low confinement ratio (ratio of particle to cylinder radius between 0.03 and 0.05) and observed particle trains above {Reynolds number $\re = 2wu_\text{max}/\nu \approx 100$.
Here, $2w$ is channel width,  $u_\text{max}$ maximum flow velocity, and $\nu$ kinematic viscosity.}
With increasing Reynolds number more and more particles assembled into trains.
For channels with quadratic or rectangular cross sections,
the particles focus on one of the four or two possible equilibrium positions depending on the aspect ratio of the 
cross section~\cite{prohm_feedback_2014,zhou_fundamentals_2013,dicarlo_inertial_2009}.
For such channels, typically, a mixture of staggered  and linear particle trains occur~\cite{hur_sheathless_2010}.
In staggered trains 
the particle locations alternate between both channel halves, 
whereas in linear trains all particles are located on one side (see \prettyref{fig:rigid_multi_setup}).
For both structures experiments and simulations  report a distinct axial spacing, where the spacing for linear trains is about twice the spacing of staggered trains~\cite{lee_dynamic_2010,gao_selfordered_2017,kahkeshani_preferred_2016,hu_inertial_2019}.

{Lee \textit{et al.} explained the formation of particle trains by the combination of two effects:
(i) inertial lift forces focus the particles onto their single-particle equilibrium positions and (ii) the 
viscous disturbance flow together with the imposed channel flow determine their axial separation~\cite{lee_dynamic_2010}.}
In Ref.\ \cite{humphry_axial_2010} the well-defined axial spacing  of two particles in a
cross-streamline pair is explained by the flow field around a single particle viewed in its center-of-mass frame.
There one observes two inward spiraling vortices on the opposite side of the channel, located around 4 particle radii 
ahead and 
behind the particle.
The second particle then follows the streamlines created by the first particle and spirals in
damped oscillations toward its equilibrium position \cite{schaaf_flowing_2019}.
This idea is also confirmed by analytical calculations~\cite{hood_pairwise_2018}.
As shown in Ref.\ \cite{schaaf_flowing_2019}, the stable cross-streamline pair also corresponds to fixed points in the lift force profiles of both particles.
Finally, two-dimensional simulations in Ref.\ \cite{hu_stability_2020} indicate
that above $\re=80$ particles in a staggered train %
perform stable oscillations about these
equilibrium positions.
The stability of linear trains is less clear.
Lattice-Boltz\-mann simulations performed by Kahkeshani \textit{et al.} indicate that same-streamline pairs are stable
\cite{kahkeshani_preferred_2016},
for which Ref.\ \cite{hood_pairwise_2018} provided an explanation based on the minimization of %
the kinetic energy of the fluid.
However, early experiments~\cite{lee_dynamic_2010} and 
most recent 2D simulations \cite{hu_inertial_2019} report an increase of 
particle spacing over time in agreement with our own findings~\cite{schaaf_flowing_2019}.
Some experiments report that for increasing Rey\-nolds number the axial spacing between the particles decreases~\cite{gupta_conditional_2018,gao_selfordered_2017} even to the value of cross-streamline pairs~\cite{kahkeshani_preferred_2016}.
However, other experiments report an increase of the spacing
in linear trains with increasing 
$\re$, while the spacing decreases 
in staggered trains~\cite{pan_direct_2018}.
In the 
range of $\re \approx 1$ to $4$ recent experiments observe that the spacing is independent of the flow velocity and that the smallest channel dimension determines the axial spacing between pairs~\cite{dietsche_dynamic_2019}.
Finally, simulations using a force coupling method report that trains are only stable 
up to lengths of 2 to 4 particles
depending on the confinement ratio and the particle Reynolds number~\cite{gupta_conditional_2018}.

A comparison of the different experimental results is hampered since they
use very different parameters, such as the channel Reynolds number and confinement ratio $a/w${, where $a$ is the particle radius and $w$ is the half width of the channel}.
While experiments with smaller particles can 
go to higher channel Reynolds numbers with ${\re>100}$~\cite{matas_trains_2004,gao_selfordered_2017}, experiments with larger particles typically operate at channel Rey\-nolds numbers between  $1$ and $20$ \cite{humphry_axial_2010,lee_dynamic_2010,dietsche_dynamic_2019}.
As the particle Reynolds number is 
$\re_p=(a/w)^2\re$, both experiments and simulations can 
operate at the same particle Reynolds number although the other parameters are very different.
This is especially relevant 
for $\re\approx100$,
where the secondary flow around a single larger particle ($a/w=0.4$) 
becomes so strong that it influences the particle-wall interactions~\cite{pan_direct_2018}.
A detailed 
review of the effect of particle size suggests that for a small confinement ratio $a/w\lessapprox0.1$ the particles move closer together with increasing Reynolds number~\cite{gao_selfordered_2017,gupta_conditional_2018,matas_trains_2004} while for larger particles the 
separation seems to increase~\cite{lee_dynamic_2010,pan_direct_2018}.
Experiments are often limited by the channel length {$L$}. 
Typical length-to-width ratios are $L/2w\approx 1000$. To overcome the 
limitation of the channel length, Dietsche \textit{et al.} used an oscillatory flow device which switched the direction 
of the flow such that the particles stayed in the channel 
and were not affected by the switching~\cite{dietsche_dynamic_2019}.
They confirmed the 
stability of cross-streamline pairs {with an
equilibrium axial distance of $\Delta z/a=3.8$}.  
For the same-streamline pair 
they identified a range of separations, $7.4<\Delta z/a<14$, where the particles moved together at low speed. 
However, the error bars were much larger than the measured speed values.

\subsection{Microfluidic phonons}
Toward the end of the article we will %
also analyze how perturbations of the regular structure move through a staggered  particle train.
An initial 
particle displacement triggers %
a displacement pulse, which travels through the 
staggered train  while being damped.
Due to 
the resemblance 
with acoustic phonons, 
such excitations are called microfluidic phonons~\cite{beatus_phonons_2006}.
So far, these phonons were only analyzed for flowing droplets squeezed between two parallel plates at vanishing Reynolds numbers
\cite{schiller_collective_2015,fleury_mode_2014}.
They provide an interesting  model system with non-linear and 
non-equilibrium behavior~\cite{beatus_twodimensional_2017}. 
In this 
quasi-two-dimensional geometry the interactions between the droplets are determined by dipolar interactions~\cite{beatus_phonons_2006}. 
Studies on such phonons in inertial microfluidics do not exist.
Due to the strong inertial damping such microfluidic phonons have to be triggered from outside.
Staggered particle trains are desired for applications since due to the regular and dense order they enhance particle throughput and at the same time facilitate particle sorting.
Analyzing how these staggered trains react on perturbations
might help to 
obtain a better understanding how 
they can be crystallized.

\subsection{Summary of results}
In this article we study the stability of staggered and linear particle trains using lattice-Boltzmann simulations.
First, we focus on a pair of flowing particles and analyze their dynamics when both particles are already initialized on their lateral equilibrium positions.
We confirm the stability 
of cross-streamline pairs and also show how they contract or expand to their equilibrium axial distance.
In contrast, same-streamline pairs quickly expand to a characteristic separation but even at long times 
slowly drift apart.
However, we are able to reproduce the distribution of  particle distances measured in experiments \cite{lee_dynamic_2010}.
Then, we extend our analysis 
to particle trains. 
We thoroughly analyze how a staggered train initialized with an axial particle spacing larger than the equilibrium distance contracts non-uniformly. 
We speculate about a possible realization and consider a defect in the train.
In contrast, particles in linear trains slowly drift apart non-uniformly, and we are able to reproduce the statistics of particle distances observed in experiments \cite{kahkeshani_preferred_2016}.
Finally, the damped displacement pulse traveling through a staggered train is presented and how a defect strongly damps its propagation.
The article is organized as follows. 
In Sect.\ 2 we explain the microfluidic setup of our system, describe 
the implementation of the lattice-Boltzmann method, and how we couple the particles to the fluid. 
In Sect. 3, we present the results for the stability of staggered and linear particle trains including particle pairs and discuss the phononic displacement pulse as well as the influence of defects.
We summarize and close with final remarks in Sect.\ 4.

\section{System and Methods}

\subsection{Microfluidic setup in the simulations}

The microfluidic setup consists of multiple particles immersed in
a Newtonian fluid with density $\rho$ and kinematic viscosity $\nu$, which flows through a microchannel (see Fig.\ \ref{fig:rigid_multi_setup}).
The channel is of length $L$ 
and has a rectangular cross section with width $2w$ and height $2h$.
To ensure that the particle dynamics is limited to the $x\mhyphen z$ plane, spanned by the short axis $x$ of the cross section and the channel direction $z$, we set the aspect ratio to $w/h=0.5$.
Thus,
only two equilibrium positions along the short 
axis exists~\cite{zhou_fundamentals_2013, prohm_feedback_2014}.
To 
realize the Poiseuille flow 
through a rectangular channel~\cite{bruus_theoretical_2008} in our lattice-Boltzmann simulations (Sect.\ \ref{sec:sim_method}), 
we drive the fluid with a  constant body force, which stands for the pressure gradient.

We 
quantify the influence of 
fluid inertia by the channel Reynolds number $\re=2wu_\text{max}/\nu$, where $u_\text{max}$ is the maximum flow velocity 
in the channel center.
In Sects.\ \ref{sec:stability_pairs} and \ref{sec:stability_trains} we restrict ourselves to the Reynolds number $\re=20$.
This corresponds to a typical experimental setup with a
channel width of $2w = \SI{25}{\micro m}$, $\nu = \SI{1e-6}{m^2/s}$ for water, and $u_\text{max} = \SI{0.8}{m/s}$ 
\cite{humphry_axial_2010}.
In Sect.\ \ref{sec:phonons} we vary the Reynolds number between 5 and 100.
In the following, we
distinguish between the axial direction along the flow ($z$ axis) and the lateral direction perpendicular to the flow ($x$ axis).
We place $N$ neutrally buoyant particle with 
radius $a$ in the channel flow.
The particles are initialized either in a staggered 
or a linear particle train (\prettyref{fig:rigid_multi_setup}).
If not stated otherwise, the particles are initialized on
the equilibrium position of single particles on the $x$ axis.
The initial axial spacing $\Delta z_0$ of two neighboring particles
in a staggered configuration differs from the observed equilibrium value
and the initial spacing in a same-streamline configuration is always chosen smaller than the limiting distance, we observe at long times.
To analyze the stability of particle pairs and trains, the
channel length was always chosen sufficiently long 
[at least $L \approx (N+1)\cdot10\,a$]
to ensure that hydrodynamic interactions with images from the periodic boundary conditions along the channel were not relevant.
When analyzing the microfluidic phonons, the channel length always was 
$L=N\cdot \Delta z$, so that we simulated an infinitely long 
particle train using periodic boundary conditions. 
Finally, in all our simulations the
particle radius was fixed to $a/w=0.4$ as in Ref.~\cite{schaaf_flowing_2019}.

\subsection{Simulation method}
\label{sec:sim_method}

To solve the Navier-Stokes equations, we 
performed simulations in 3D with the lattice-Boltzmann method (LBM)
using 19 different velocities vectors (D3Q19) \cite{succi_lattice_2001} and 
the Bhatnagar-Gross-Krook (BGK) collision operator \cite{bhatnagar_model_1954}. 
{We relied on the same simulation code used in our previous publication \cite{prohm_feedback_2014, schaaf_flowing_2019}.
The LBM 
determines the one-particle probability distribution $f_i(\vec x,t)$
on a cubic lattice with lattice spacing $\Delta x=1$.
In addition
to 
space and time discretization, the LBM also discretizes the possible 
velocity vectors, which are indicated by the index $i$ in $f_i(\vec x,t)$.
In the case of the velocity set D3Q19, 
19 
velocity vectors are implemented. 
During time $\Delta t$ the
particle distribution function $f_i(\vec x, t)$ evolves 
according to two alternating steps:
\begin{align}
\text{collision: }& f_i^\ast(\vec x,t)=f_i(\vec x,t)+\frac{1}{\tau}\left[f_i^\text{eq}(\vec x,t)-f_(\vec x,t)\right]\\
\text{streaming: }& f_i(\vec x+\vec c_i\Delta t,t+\Delta t) = f_i^\ast(\vec x,t) \, ,
\end{align}
where $f_i^\text{eq}$ is a second-order expansion of the Maxwell-Boltz\-mann distribution  in the mean velocity and $\tau$ is the relaxation time of the BGK model. 
For details on this method we refer the interested reader to Ref.~\cite{kruger_lattice_2017}.

One important feature of the LBM is that the viscosity is related to the collision relaxation time $\tau$~\cite{dunweg_lattice_2008},
\begin{equation}
\nu=c_s^2\Delta t\left(\tau-\frac 12\right) \, ,
\end{equation}
where $c_s^2=1/3$ is the speed of sound measured in LBM units ($\Delta x=1$, $\Delta t=1$). 
The lattice-Boltzmann method does not strictly conserve 
volume.
To ensure incompressibility of the fluid, simulations have to be performed at small Mach numbers $\mathrm{Ma}=u_\text{max}/c_s$. 
For all 
our simulations the Mach number is set to $\mathrm{Ma}\leq0.1$, which corresponds to {a maximum possible density variation of 
1\%.}
The exact procedure is explained in Ref.~\cite{schaaf_flowing_2019}.

To drive the channel flow, we apply a constant body force according to the Guo-force scheme~\cite{guo_discrete_2002}. 
At the channel walls we 
use regularized boundary conditions~\cite{latt_straight_2008}. 
The main part of our simulation code 
is provided by the Palabos project~\cite{_palabos_2013}, which we extended by implementing the flowing 
particles.
We use a resolution of 60 lattice cells along the short axis.

For 
simulating the suspended particles, we rely on the immersed-boundary method (IBM) proposed by Inamuro~\cite{inamuro_lattice_2012}. 
This scheme belongs to the class of 
immersed-boundary methods with direct forcing and was shown to capture the correct behavior of solid particles in laminar flow~\cite{derosis_comparison_2014}.
The immersed-boundary method is based on a Lagrangian grid, which exists in between the fixed Eulerian grid for the fluid.
The velocity of the fluid is interpolated 
to the particle surface, where a penalty force ensures the no-slip boundary condition.
The resulting force is interpolated back to the fluid.
This process is done iteratively, where we use the recommended five iterations. For further details on these methods we refer the reader to the original publications~\cite{inamuro_lattice_2012} and our previous work~\cite{schaaf_flowing_2019}.

\section{Results} \label{sec:results}

We now present our results starting with the stability of cross-streamline and same-streamline pairs, which we then extend to the same types of particle trains. 
Finally, we address microfluidic phonons concentrating on the damped propagation of a displacement pulse.

\subsection{Stability of particle pairs}
\label{sec:stability_pairs}

Before we consider multi-particle trains, we first analyze the long-time behavior and
stability of a pair of rigid particles where both particles are initialized at their lateral single-particle equilibrium positions. 
We focus on the axial distance and look for stable axial configurations.
We already analyzed the dynamics of 
a pair of particles interacting by inertial lift forces and briefly summarize here the results relevant for this work~\cite{schaaf_flowing_2019}.

Analyzing 
two-particle trajectories with different initial positions, we only found 
one 
class of bound
trajectories, where two particles on opposing sides of the channel (cross-streamline configuration) performed 
damped oscillations toward their equilibrium positions.
All other types of trajectories were unbound; the axial particle distance increased while
the two particles moved individually toward their single-particle equilibrium positions.
In addition, an analysis of the two-particle lift force profiles showed that cross-streamline configurations are stable and thereby confirmed observations 
by other groups~\cite{humphry_axial_2010,kahkeshani_preferred_2016,hood_pairwise_2018}.
For the same-streamline configuration we 
could not identify a fixed point 
with a small axial particle distance, which is reported in experiments at larger Reynolds numbers $\re=60$ and $90$ \cite{kahkeshani_preferred_2016}.
Instead, we observed that the particles moved apart until their interaction was vanishingly small
close to the experimentally reported larger spacing \cite{kahkeshani_preferred_2016,dietsche_dynamic_2019}.
Hence, we concluded that same-streamline 
pairs are only one-sided stable: the particles repel after being pushed together, while, when moved apart, they keep their distance due to the missing attraction.
A similar observation was reported 
in one of the earliest works on particle trains in inertial microfluidics~\cite{lee_dynamic_2010}.

In the following, we  focus on the case where both particles already occupy
their equilibrium lateral positions.
Thus, they stream with the same velocity.

\subsubsection{Cross-streamline particle pairs}
\begin{figure}%
	\centering
	\includetikz{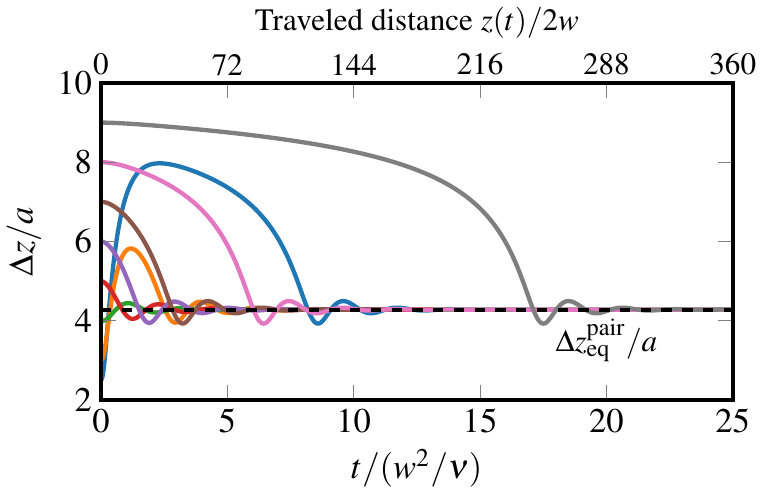}
	\caption{
		Axial distance $\Delta z/a$ as a function of time for a particle pair in %
		cross-streamline configuration.
		Both particles are initialized at single-particle equilibrium positions,
		$x=\pm x_\text{eq}^\text{single}$,
		and different initial axial distances $\Delta z_0$ are chosen.
		The dashed line indicates the equilibrium axial distance
		 $\Delta z_\text{eq}^\text{pair}/a\approx4.2$.
		The upper axis 
		indicates the traveled distance of the center-of-mass of the pair along the channel for $\Delta z_0/a = 8$.
	}
	\label{fig:rigid_pair_contraction_distance}
\end{figure}

\begin{figure}%
	\centering
	\includetikz{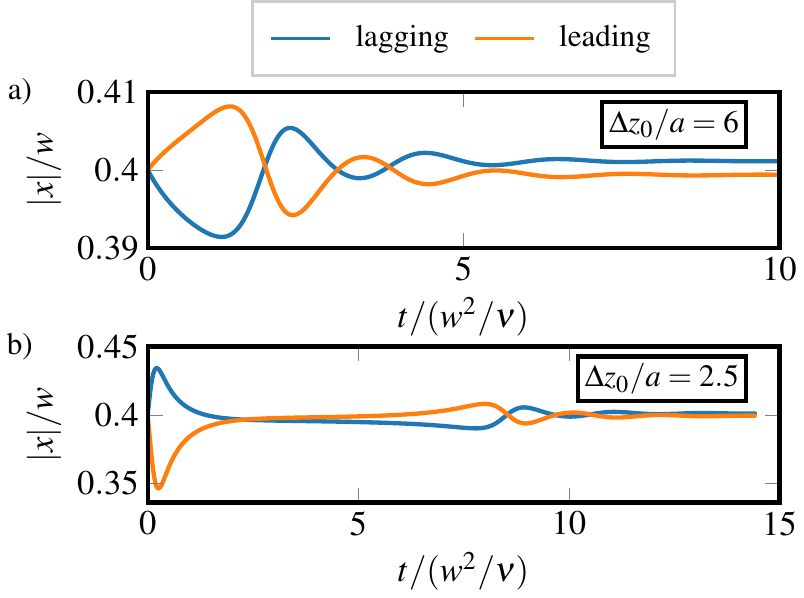}
	\caption{
        Leading and lagging particles in a cross-streamline configuration.
        Distance from the channel centerline $|x|/w$ as a function of time.
		Both particles are initialized at $x=\pm x_\text{eq}^\text{single}$. 
		(a) Initial distance $\Delta z_0 = 6a  > \Delta z_\text{eq}^\text{pair}$ and 
		(b) $\Delta z_0 = 2.5 a  < \Delta z_\text{eq}^\text{pair}$.
}	
	\label{fig:rigid_pair_contraction_lateral}
\end{figure}

In our previous work we found that 
particles in all analyzed bound trajectories reach
the same value for their axial distance, $\Delta z_\text{eq}^\text{pair}=4.2\,a$, independent of the initial positions{~\cite{schaaf_flowing_2019}}.
In order to observe the damped oscillation, 
leading and lagging particles in flow were initialized with a similar lateral position 
$x_\text{lag}\approx -x_\text{lead}$.
Here, we 
analyze 
the situation where both particles are initialized on the single-particle equilibrium positions at $\pm x_\text{eq}/w\approx \pm0.4$ but with a distance $\Delta z_0 \ne \Delta z_\text{eq}^\text{pair}$.
In \prettyref{fig:rigid_pair_contraction_distance} we
vary the initial distance $\Delta z_0$ from $2.5\,a$ to $9\,a$ and plot the respective time course of the axial distance.
In all cases the particles 
reach $\Delta z_\text{eq}^\text{pair}\approx4.2$, even when the initial axial distance 
is as large as $9\,a$. 
In the graph we
also added the traveled distance of the particle pair for $\Delta z_0/a = 8$.
This shows that the particle pair relaxes to its equilibrium configuration on distances much shorter than typical channel lengths of the order of $L/2w\approx 1000$ \cite{lee_dynamic_2010,gao_selfordered_2017}.

In \prettyref{fig:rigid_pair_contraction_lateral} we show  the time course of the lateral coordinates of the leading and lagging particles.
When the initial axial distance $\Delta z_0$ is larger than the equilibrium value
[\prettyref{fig:rigid_pair_contraction_lateral}\,(a)], the lateral positions hardly change.
But this is sufficient to let them approach each other.
Only when the particles are initialized closer together [\prettyref{fig:rigid_pair_contraction_lateral}\,(b)], do the lateral position change noticeably by ca. 10\%. 
This then initiates the 
initial rapid increase of the axial spacing due to the different flow velocities [blue line in \prettyref{fig:rigid_pair_contraction_distance} with $\Delta z_0/a=2.5$] followed by a slow relaxation back to the equilibrium value.
Thus, our study also shows how
particle pairs 
relax toward their preferred distance even when starting at
their lateral equilibrium positions.
We note that in the final configuration the leading particle 
is located slightly closer toward the channel center but moves with the same velocity.
However, the difference is smaller than a lattice unit.

\subsubsection{Same-streamline particle pairs}
\begin{figure}[bt]
	\centering
	\includetikz{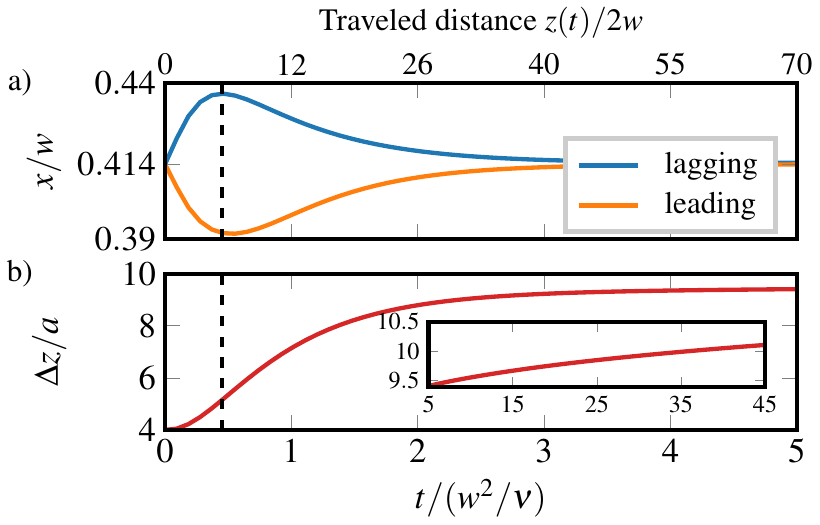}
	\caption{
		(a)~Lateral positions and (b)~axial distance as a function of time for a same-streamline particle pair. 
		The dashed line localizes the maximum lateral displacement.
        Inset: At larger times still a slow but steady increase of $\Delta z$ is observable.
		The upper axis 
		indicates the traveled distance of the center-of-mass 
		of the pair along the channel.		
		At $t=45 w^2/\nu$ the particles have moved a distance of $z/2w=650$.
		}
	\label{fig:rigid_pair_spreading_distance_lateral}
\end{figure}

In Ref.\ \cite{schaaf_flowing_2019} we already observed that particles
initialized on the same streamline with an initial distance of $\Delta z_0/a=5$ slowly drift apart, even when 
positioned on the single-particle equilibrium position.
We explained this behavior with an asymmetry in the two 
corresponding lift force profiles: 
the leading particle is pushed toward the channel center 
while the lagging particle toward the walls such that they move apart. This behavior did not change for larger $\Delta z_0$.
Hence, 
our previous simulations indicate that same-streamline pairs are not stable.

We now analyze 
this behavior in more detail in \prettyref{fig:rigid_pair_spreading_distance_lateral}.
Again, we initialize the 
particles at the lateral equilibrium position and with an 
axial spacing
equal to the
axial distance $\Delta z_\text{eq}^\text{pair}$ of the cross-streamline pairs.
The
leading particle is noticeably and rapidly pushed toward the channel center while the lagging particle  
moves outwards [\prettyref{fig:rigid_pair_spreading_distance_lateral}\,(a)].
This drives the particles apart, which is visible by the
rapid increase of the axial distance 
in \prettyref{fig:rigid_pair_spreading_distance_lateral}\,(b).
Beyond $\Delta z/a\approx 5$, the particles slowly relax toward their equilibrium lateral position.
However, even at large times, where
both particles should move with the same speed, we still observe a slow but steady
increase of the particle spacing [\prettyref{fig:rigid_pair_spreading_distance_lateral}\,(inset)].
In experiments such a drift might be difficult to measure. 
According to our simulations, beyond 
$t= 5 w^2/\nu$ the distance $\Delta z$ increases by only 6\% while the particle pair travels a distance of $600 \times 2w$.
This is of the order of channel lengths used in experiments. 
Taking typical values of $L=\SI{5}{cm}$ and $2w =  \SI{50}{\micro m}$,
we obtain $L/(2w)\approx 1000$ \cite{gao_selfordered_2017}.

\begin{figure}[bt]
	\centering
	\includetikz{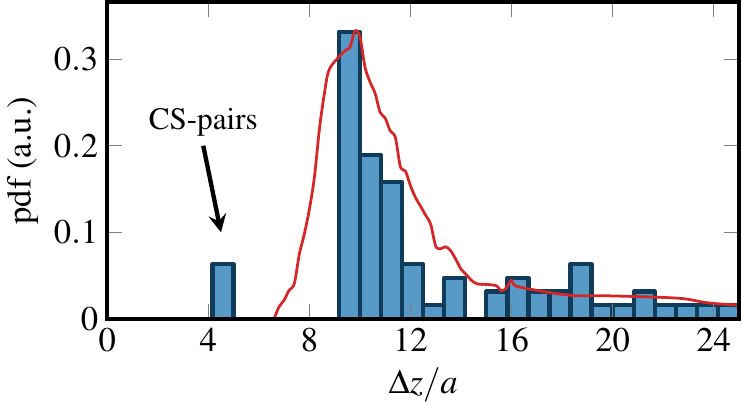}
	\caption
	{Histogram of particles distances 
    for randomly initialized pairs after 
    the center-of-mass has traveled a distance of
    $z/2w=1000$.		
	The small peak at $\Delta z/a=4$ corresponds to cross-streamline pairs, which formed despite the fact that all particles 
	were initialized on the same channel side.
	The red line shows data from experiments 
	for pairs of particles with the same confinement ratio($a/w=0.4$) and the same traveled 
	distance (\SI{2.5}{cm})~\cite{lee_dynamic_2010}.
	We rescaled the experimental data by a factor of \num{0.4} to match the height of the peak.	
	}
	\label{fig:rigid_multi_pair_distance_histogram}
\end{figure}

Experiments typically report a distance of about twice the equilibrium spacing of cross-streamline pairs, which would be $\Delta z/a\approx 8.4$ in our case~\cite{kahkeshani_preferred_2016}.
Our simulations indicate that the particles go to a somewhat larger spacing.
In experiments the starting conditions are not as well-defined as in our case.
To reproduce the experimental statistics for 
particle distances observed in Ref.\ \cite{lee_dynamic_2010}, we initialize 76 different pair configurations, where both particles are randomly placed in the upper channel half with an initial distance 
$5a < \Delta z_0< L/2$.
In \prettyref{fig:rigid_multi_pair_distance_histogram} we
plot the distribution of 
particle distances after the particles have traveled a distance of $1000 \times 2w$, which is a typical value in experiments as mentioned above.
The simulated distribution matches well with experiments for the same particle size with a peak at $\Delta z/a=9.8$ \cite{lee_dynamic_2010}. 
The authors do not specify the channel Reynolds number explicitly, but the particle Reynolds number $\re_p = \re (a/w)^2 = 2.4$
is similar to the value of $3.2$ used for this work.

To summarize, while we obtain good agreement
with early experiments~\cite{lee_dynamic_2010},
we could neither 
identify an additional stable equilibrium distance
for higher Reynolds numbers~\cite{kahkeshani_preferred_2016} nor reproduce an attractive 
interaction of particles in
same-streamline pairs~\cite{dietsche_dynamic_2019}.

\subsection{Stability of particle trains}
\label{sec:stability_trains}
Based on the insights we gathered from the behavior of particle pairs, we continue by analyzing multi-particle trains and begin with staggered particle trains.

\subsubsection{Staggered particle trains}

\begin{figure}[bt]
	\centering
	\includetikz{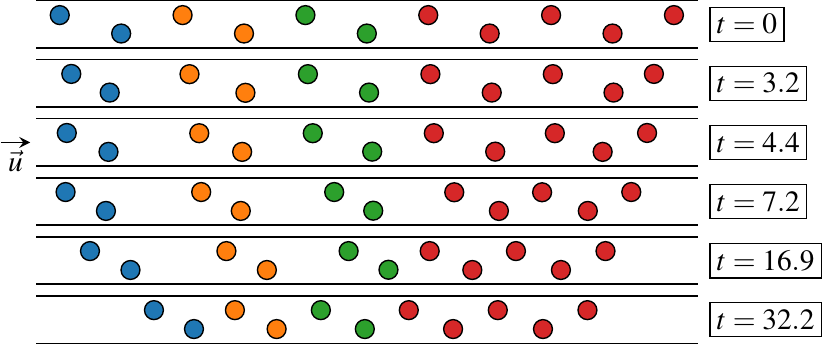}
	\caption{
Snapshots of the contraction process of a staggered particle train at different times given in units of $w^2 / \nu$. 
At $t=0$ all particles are initialized with a nearest-neighbor distance $\Delta z_0 =6.5 a =1.5\Delta z_\text{eq}^\text{pair}$.
{The color of the trailing pairs corresponds to the lines in \prettyref{fig:rigid_zigzag_contraction_distance_time}\,(b).}
	}
	\label{fig:rigid_zigzag_contraction_pictures}
\end{figure}

\begin{figure}[bt]
	\centering
	\includetikz{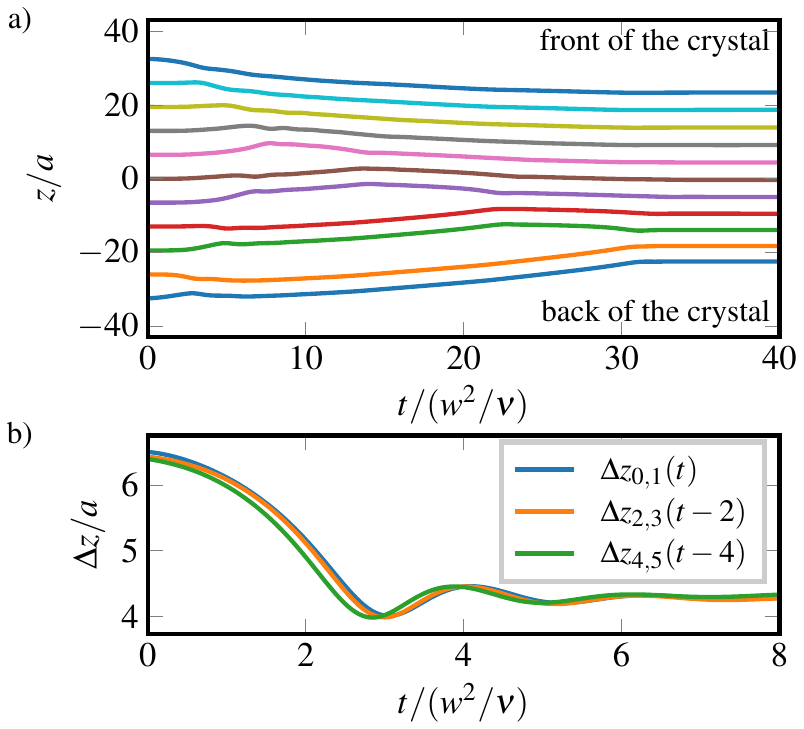}
	\caption{
		(a) Axial positions of all the 
		particles 
		in a staggered particle train
		plotted versus time. 
		The positions are given in
		the center-of-mass frame of the train.
		(b) Axial particle distances
		of the trailing particle
        pairs as a function of time. 
		The curves of
        the second and third pair are shifted such that 
        they fall onto each other.
		The initial axial distance is $\Delta z_0=6.5\,a\approx 1.5\Delta z_\text{eq}^\text{pair}$
	}
	\label{fig:rigid_zigzag_contraction_distance_time}
\end{figure}

When we start the simulations with arbitrarily placed particles, we obtain staggered particle trains with defects in it.
To concentrate first on the ideal case, in
the following we analyze how 
an expanded staggered particle train contracts toward its equilibrium configuration.
For this we consider 11 particles, which we initialize on their single-particle position 
at
$\pm x_\text{eq}$ with an axial distance of $\Delta z_0=6.5\,a\approx 1.5\Delta z_\text{eq}^\text{pair}$.

As expected from the analysis of the cross-streamline pairs, the axial distances between the particles decrease in time.
However, as Figs.~\ref{fig:rigid_zigzag_contraction_pictures}, \ref{fig:rigid_zigzag_contraction_distance_time}(a), and video 1 in the supplemental material demonstrate, the contraction does not occur uniformly but rather through the formation of particle pairs. The
contraction starts in the front and 
back of the train. 
We observe that initially only the leading and trailing pairs 
contract, 
whereby mainly the leading particle of the pair moves backwards toward the lagging particle [\prettyref{fig:rigid_zigzag_contraction_pictures}\,($t=3.2$)].
While the pairs contract, they slow down. 
The leading pair stays connected to the staggered train but the last pair
separates from the rest of the train due to its reduced velocity (see below).
This triggers the contraction of the next pair and then a third pair so that at $t=7.2$ three individual pairs in the back of the train exist.
The contraction of these pairs 
always occurs in the same manner. 
In \prettyref{fig:rigid_zigzag_contraction_distance_time}\,(b) we plot their particle distances versus time and have shifted the curves by the time the previous pair needed to contract and separate.
Then, all three curves fall onto each other.
The particle pair in the front of the staggered train also slows down.
The next particle in line catches up so that a three-particle cluster exists at $t=4.4$. 
This cluster slows down further and the next two particles can catch up.
Ultimately, at $t=7.2$ a contracted five-particle cluster exists 
followed by the 3 trailing pairs.
The larger cluster slows down (see below) so that the three pairs can catch up one by one ($t=16.9$) and, finally, at $t=32.2$
the staggered train 
has reached its equilibrium configuration.

\begin{figure}[bt]
	\centering
	\includetikz{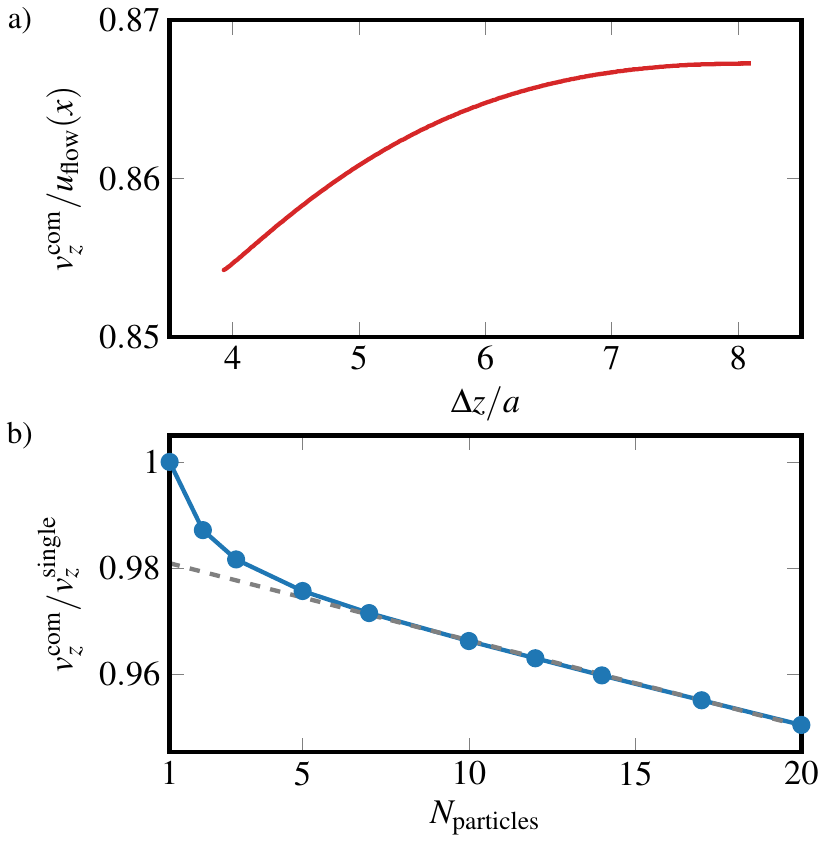}
	\caption{
		(a)~Axial center-of-mass velocity for a cross-streamline pair as a function of the particle distance. 
		The velocity is plotted in units of the fluid flow velocity at the lateral position of the particle pair.
		The data were extracted from
		the gray curve in \prettyref{fig:rigid_pair_contraction_distance}, which starts at $\Delta z_0=9 a$.
		(b)~Axial center-of-mass velocity for a staggered particle train as function of the number of particles in the train.
		The velocity is plotted in units of the single-particle velocity.
		The gray dashed line is a linear fit for range with $N>5$.
	}
	\label{fig:rigid_multi_cluster_speed}
\end{figure}

For 
the contraction of the particle train, two mechanisms are relevant. 
They are related to viscous drag reduction of clusters of particles compared to a single particle and when the clusters are more compact \cite{reichert_circling_2004,janssen_collective_2012,beatus_twodimensional_2017}.
In our case, this means the resistance to an imposed Poiseuille flow is reduced and therefore the clusters slow down relative to the flow.
Thus,
a pair of particles slows down when the axial distance decreases and
the center-of-mass velocity also decreases for larger staggered trains.
We discuss this in detail in
\prettyref{fig:rigid_multi_cluster_speed}.
In graph (a) we plot
the center-of-mass velocity of a  
cross-streamline pair as a function of the axial particle distance.
Although the 
decrease of the velocity with $\Delta z$ is very small, it quantitatively agrees with the 
reported results of 
simulations for a pair of rigid particles ($a/w=0.8$) moving on
the centerline of a confined flow at $\re\ll1$~\cite{janssen_collective_2012}.
In plot (b) we observe for
staggered particle trains
that the center-of-mass velocity monotonically decreases with increasing number of particles.
For $N>5$ this decrease is linear. 
The difference in velocities of
a cluster consisting of 20 particles and a single particle is about \SI{5}{\percent}. A very similar 
dependence on the particle number was reported for simulations of a chain of particles 
driven by an applied force along a ring  
in a bulk fluid~\cite{reichert_circling_2004}. 
However, in 
this situation the variation of the velocity is more pronounced (around \SI{50}{\percent}). 
The same type of collective drag reduction was also reported by Beatus \textit{et al.}~\cite{beatus_phonons_2006} for a linear chain of droplet disks in a quasi-2D flow.
The authors
named this 
observation the \textit{peloton effect}, in analogy to the reduced drag of a 
group of closely riding cyclists.

\begin{figure}[bt]
	\includetikz{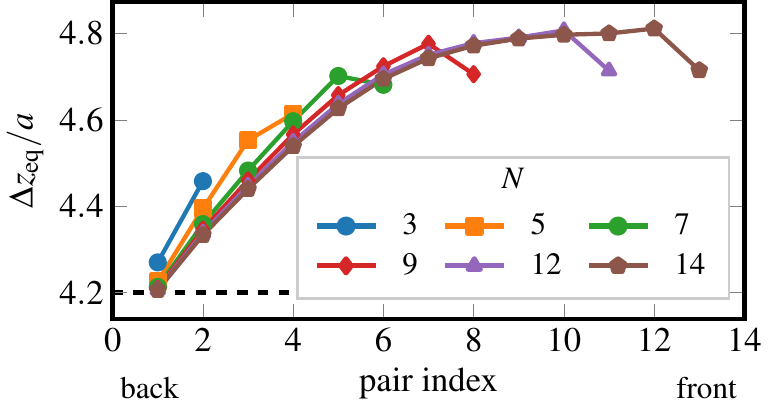}
	\caption{
		Axial equilibrium distance 
		of neighboring particles in a staggered train as a function of 
		the pair index,
        which increases from the back to the front of the {train}.
        The dashed line indicates the particle distance of a single pair, $\Delta z_{\mathrm{eq}^{\mathrm{pair}}}/a = 4.2$.
	}
	\label{fig:rigid_zigzag_delta_z}
\end{figure}

We finish with a final observation.
For cross-streamline pairs we found that all particles 
relax toward the same value $\Delta z/a=4.2$ for the axial distance.
In contrast, as \prettyref{fig:rigid_zigzag_delta_z} shows the axial spacing of neighboring particles in a staggered train is non-uniform. 
It
increases when one moves forward in the train.
For 
trains with more than nine
particles the axial distance saturates at a value of $\Delta z/a=4.8$, which is about 15\% larger than the distance of a single pair.
Finally, we observe that the distance of the leading pair is slightly reduced for 
trains consisting of 
seven particles or more.

\subsubsection{Linear particle trains}
\begin{figure}[bt]
	\centering
	\includetikz{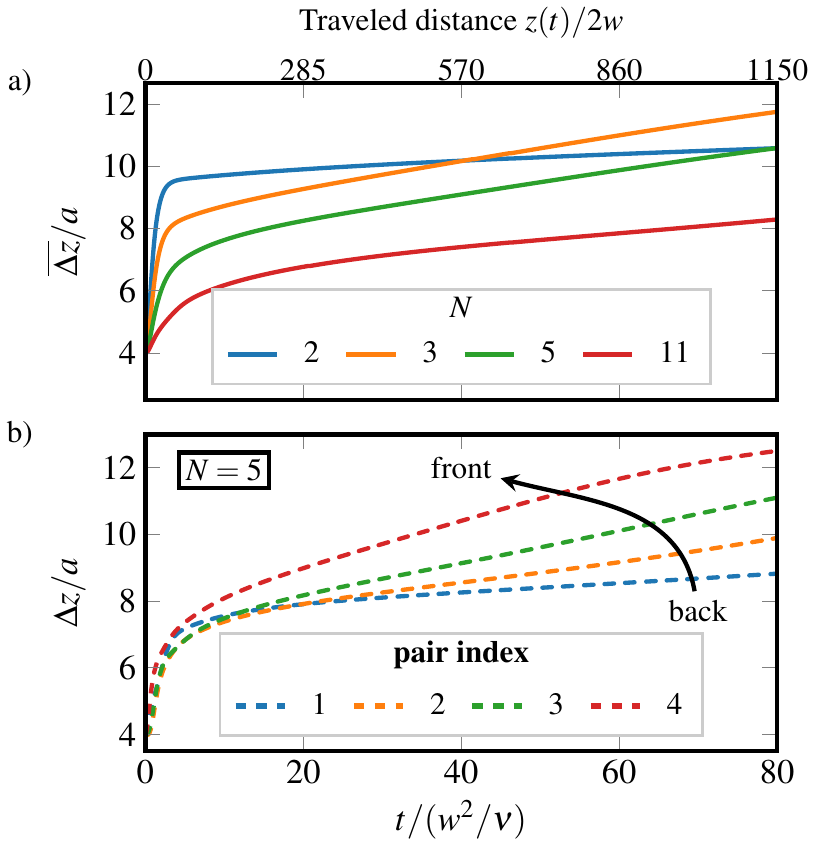}
	\caption{
		(a)~Mean axial particle distance as a function of time for linear trains with different numbers of particles. 
		(b)~Axial distance 
		of neighboring particles within a linear train consisting of 5 particles. 
		From the back to the front the pairs are indexed by 1 to 4.
	}
	\label{fig:rigid_multi_mean_distance}
\end{figure}

\begin{figure}
	\centering
	\includetikz{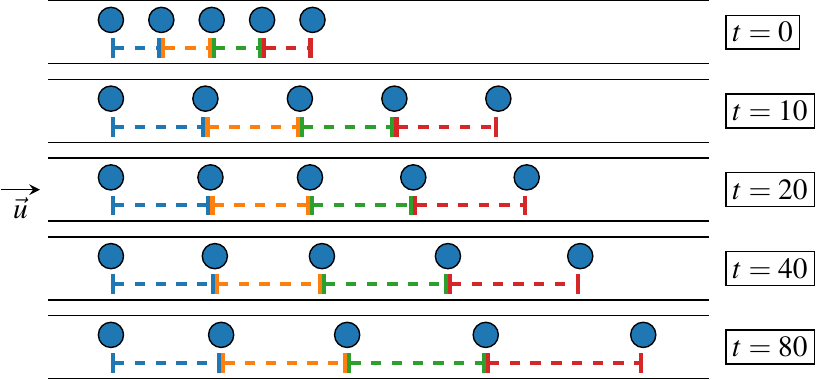}
	\caption{%
		{Snapshots  
		of an expanding linear particle train at different times given in units of $w^2 / \nu$. 
		At $t=0$ all particles are initialized 
		with the axial equilibrium distance 
		of
		cross-streamline pairs.
	   The dashed lines correspond to $\Delta z$ plotted in \prettyref{fig:rigid_multi_mean_distance}\,(b).}
	   }
	\label{fig:linear_train_snapshot}
\end{figure}

Linear particle trains flowing on the same streamline have been observed both
in experiments~\cite{gao_selfordered_2017,kahkeshani_preferred_2016} and simulations~\cite{gupta_conditional_2018}.
Typically, they have an axial particle spacing about twice the distance 
measured for staggered trains.
So
far, in Sect.\ \ref{sec:stability_pairs} we found
that the axial distance of a same-streamline particle pair steadily increases in time.
In the following we analyze the stability of  
longer particle trains and check if multi-particle interactions can 
stabilize them.

We initialize the particles in a linear train at the 
lateral equilibrium positions of single particles and choose an initial axial distance of $\Delta z_0/a=4$ between neighboring particles. Figure\ \ref{fig:rigid_multi_mean_distance}\,(a) shows that
the mean axial distance for linear trains of different sizes
increases monotonically in time.
However, while for $N=2$ the axial distance hardly changes after
reaching a distance of $\Delta z/a\approx 10$, the mean distance of trains with $N>2$ increases visibly
and 
the increase is slower for longer trains.

The reason is that the expansion of the linear train is non-uniform as we show in \prettyref{fig:rigid_multi_mean_distance}\,(b), where we plot the axial distance of neighboring particles for a train with five particles. 
{Snapshots of the expanding train are
presented
in \prettyref{fig:linear_train_snapshot}.
}
The expansion for five and 11-particle trains is also visualized in videos 2 and 3 of the supplemental material.
After an initial fast expansion to $\Delta z/a\approx 8$, which is about twice the distance in a cross-streamline pair, always the leading particle is carried away by the imposed flow while the particle train behind it moves more slowly. 
This creates a particle train where the particle distance at one instant in time increases from the back (pair index 1) to the front (index 4). 
In our simulations all particles have traveled a distance of more than $1000w$ as the upper axis of the plot shows. 
Thus, the steady increase of $\Delta z$ is very slow and it
might not be possible to observe this in a typical experiment. 
However, our data show that the final configuration in the simulations is not a stable configuration.

The separation of the leading particle from the rest of the train
was also reported in simulations by Gupta \textit{et al.}~\cite{gupta_conditional_2018}. 
They also mention stable trains 
up to a certain cluster size,
an
effect the authors name \textit{conditional stability}.
Their critical cluster size
depends on the confinement ratio $a/w$ and the particle Reynolds number. 
In their analysis 
the authors focused on confinement ratios 
$a/w=\numrange[range-phrase=-]{0.08}{0.14}$, which is much smaller than in our work. However, their results indicate that for larger particles the critical cluster size reduces to $N=2$.

\begin{figure}[bt]
	\centering
	\includetikz{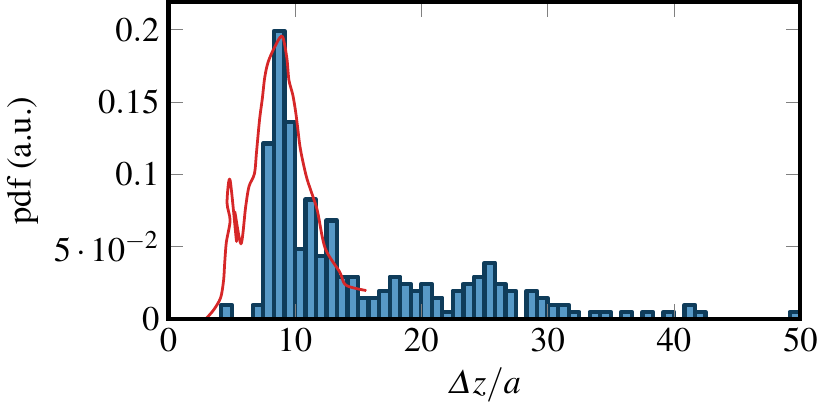}
	\caption{
		Histogram of particles distances for 4, 6, or 11 particles randomly initialized in the upper channel half after they 
		have traveled at least a distance of $160w$ along the channel. The small peak at $\Delta z/a=4$ corresponds 
		to cross-streamline pairs, which formed despite the fact that all particles were initialized on the same channel side.
		The red line shows data from experiments by Kahkeshani \emph{et al.} \cite{kahkeshani_preferred_2016} 
		for slightly smaller particles with $a/w=0.34$.
		We rescaled the experimental data by a factor of \num{0.7} to match the height of the peak.	
	}
	\label{fig:rigid_multi_train_distance_histogram}
\end{figure}

Finally, we calculate the statistics 
of the  particle spacing 
in linear trains (\prettyref{fig:rigid_multi_train_distance_histogram}).
For this we randomly place 4, 6, 
or 11 particles in the upper channel half
and ensure that there is no overlap between the particles. 
The channel length 
is chosen such that the volume fraction 
is fixed at $\varphi = 0.004$ 
as in Ref.~\cite{kahkeshani_preferred_2016}. 
After the particles traveled at least a distance of 
$160\,w$, we measure the distance to the 
nearest-neighbor particles. 
The recorded statistics shows
good agreement with 
experimental data measured for slightly smaller particles with $a/w=0.34$ compared to our particles with $a/w=0.4$ (\prettyref{fig:rigid_multi_train_distance_histogram}).
Again, we observe a small peak at $\Delta z/a=4.2$ which, in our case, corresponds to particles which move to the lower channel half.
{This first peak is much smaller than in the experiments. This is most probably due to how the linear trains are set up.
In the experiment the same-streamline trains are created 
with two inlets which form a Y-shaped channel, where the particles enter via one inlet. It sounds plausible that 
with this method cross-streamline particle pairs are more easily formed compared to our simulations, where we initialize 
the particles already on the same streamline.
}

\subsubsection{Staggered particle train with defect}

\begin{figure}[bt]
	\centering
	\includetikz{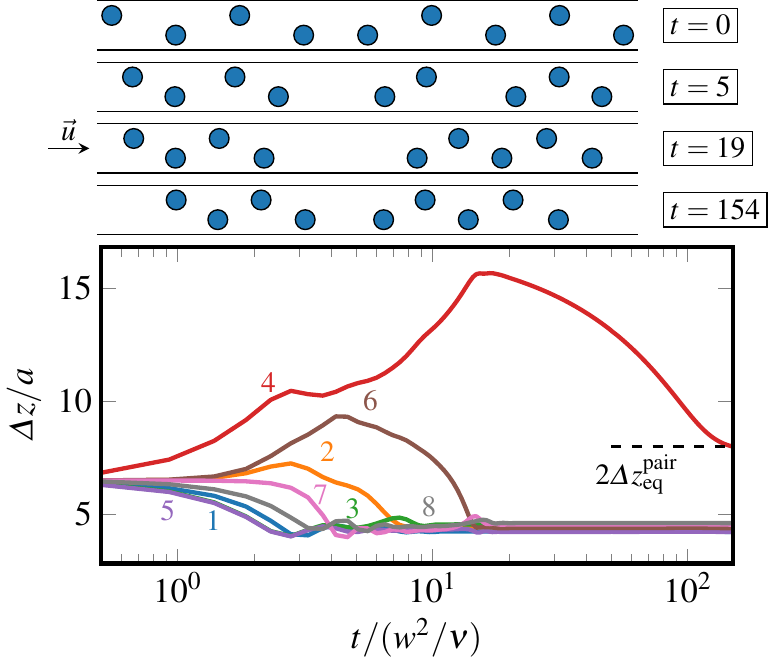}
	\caption{
		Staggered particle train with defect. 
		Top: {Snapshots at different times $t$ of the non-uniform contraction from the initial ($t=0$)
		 toward the final $(t=154)$ configuration.}
		Bottom:~Distances between neighboring particles plotted versus time. 
		The numbers indicate the pair index increasing from the back to the front.
        The final particle distance of the defect is $2\Delta z_{\mathrm{eq}}^{\mathrm{pair}}$ as indicated by the dashed line.
}
	\label{fig:rigid_mixed_structure}
\end{figure}

Besides the pure cross-streamline and same-streamline particle trains we also initialized 
a staggered train with a single defect.
The results for the temporal evolution of the particle distances and the final configuration are presented in \prettyref{fig:rigid_mixed_structure}.

To create a defect, the fourth and fifth particles are placed on the same channel side so that
two staggered trains exist, which consist
of four and five particles, respectively. 
The trains first contract individually while they drift apart from each other as the increasing distance of particle 4 and 5
indicates (red line in \prettyref{fig:rigid_mixed_structure}, see also video 4).
Only after the 
two trains have reached their equilibrium configuration, we observe that the lagging 
four-particle train catches up to the slightly slower 
train with five
particles  [see Fig.\ \ref{fig:rigid_multi_cluster_speed}(b)]. 
In the final state the particle distance of the defect is
about twice the equilibrium particle distance of 
cross-streamline pairs 
similar to observations in~\cite{hur_sheathless_2010,edd_controlled_2008}.
We note that this distance is not governed by 
any attractive interaction between the two particles
but solely due to the fact that the larger leading train
moves slower than the smaller 
trailing train. 
Indeed, when we swap the two trains such 
that the smaller 
one is leading, the two 
trains slowly drift apart in time.

\subsection{Microfluidic phonons}
\label{sec:phonons}

Regular structures such as the staggered particle trains can be perturbed and thereby show propagating phononic excitations or microfluidic phonons.
To study them in more detail, we analyze how a cross-streamline train reacts to a perturbation of a single particle position. 
Video 5 in the supplemental material shows the resulting damped pulse propagation.
We fit a train of 12 particles into a channel and adjust its length such that periodic boundary conditions generate an infinitely extended staggered train.

In \prettyref{fig:rigid_multi_phonon_oscillations}(a) we show the staggered train, where we initialized the
12 particles with an axial spacing of $\Delta z/a=4.2$, which corresponds to the equilibrium distance of 
an isolated particle pair, and with lateral equilibrium positions at
$\pm x_\text{eq}/w = \pm 0.4$.
To perturb the system, we move one particle 
inwards toward the channel center as indicated. 
Thus, it moves faster than the train and approaches the neighboring particle in front.
Figure\ \ref{fig:rigid_multi_phonon_oscillations}(b) quantifies the reaction of all the particles by plotting their displacements 
$\Delta |x(t)|=|x(t)| - x_\text{eq}$ from the equilibrium position where $\Delta |x| < 0$ means motion toward the channel 
center and $\Delta |x| > 0$ toward the channel wall. 
While approaching the neighboring particle, the first particle is pushed back to its  equilibrium position and thereby pulls the neighboring particle from the opposite channel side toward the center. 
Thus, the whole process repeats such that a displacement pulse travels through the staggered {train} as illustrated by \prettyref{fig:rigid_multi_phonon_oscillations}(b). 
The particle motion is strongly damped since the inertial lift force pushes the particles back toward their equilibrium positions. 
Thus, the initially displaced particle (pink curve) overshoots only by a small distance and then relaxes toward its equilibrium position. 
Also, the propagating displacement pulse is exponentially damped (dashed line in \prettyref{fig:rigid_multi_phonon_oscillations}) with a damping rate $\gamma$, which we discuss in more detail further below. 
We observe that individual particles in a train return much faster to their equilibrium positions compared to isolated particles due to the coupling to neighboring particles, while the relaxation time of the whole pulse, $\gamma^{-1}$, is roughly twice as big.
Below, we will also discuss in more detail the velocity of the displacement pulse.

\begin{figure}[bt]
	\centering
	\includetikz{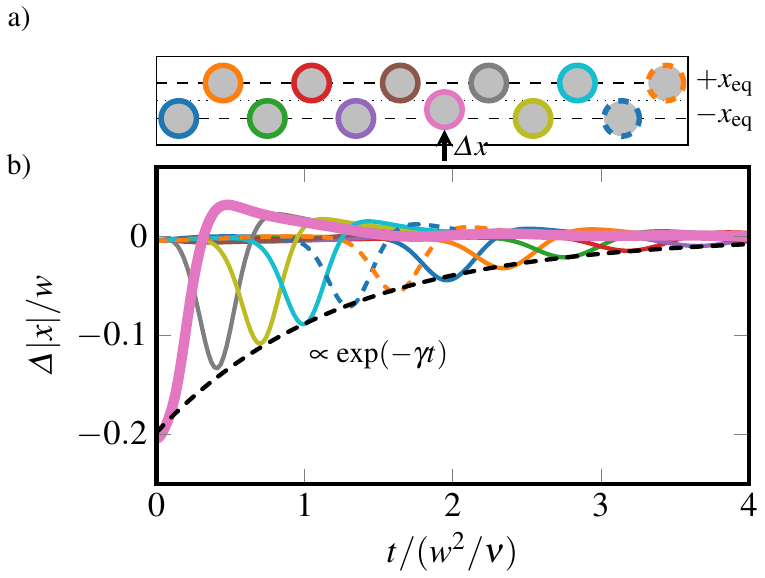}
	\caption{
	(a)~Staggered particle train with an initial displacement of the seventh particle counted from the end (pink)	and equilibrium axial particle distance is $\Delta z /a = 4.2$.
	(b)~Lateral particle displacement from the equilibrium position, $\Delta |x(t)|=|x(t)| - x_\text{eq}$, plotted versus time for all the particles. 
	The color coding is the same as in (a). 
	Here $\Delta |x| < 0$ means motion toward the channel center and $\Delta |x| > 0$ toward the channel wall. 
	The exponential decay of the pulse height with time is indicated (dashed line) and $\gamma$ is damping rate. 
	The Reynolds number is $\re=25$.
}
	\label{fig:rigid_multi_phonon_oscillations}
\end{figure}

\begin{figure}%
	\centering
	\includetikz{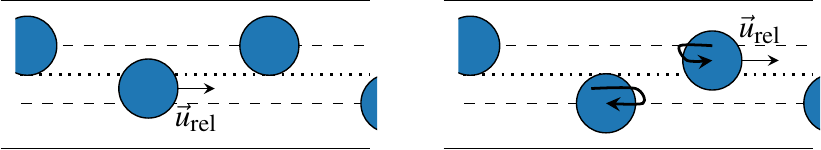}
	\caption{
	The swapping mechanism from Ref.~\cite{schaaf_flowing_2019}	explains how the displacement pulse is passed from the lagging to the leading particle. 
	The curved arrows indicate the 
	particle trajectories during swapping.
	The lateral equilibrium position of the particles and the channel center are marked by the dashed and dotted lines, respectively.
    The swapping mechanism is also visualized in video 6 of the
supplemental material.
	}
	\label{fig:rigid_multi_phonon_swapping}
\end{figure}

The mechanism for the propagating displacement pulse
can be explained as a 
sequence of swapping trajectories similar to the one we discussed in a previous work~\cite{schaaf_flowing_2019}. 
As indicated in Fig.\ \ref{fig:rigid_multi_phonon_swapping}, the displaced particle approaches the next particle in line, and they swap their lateral positions such that $x_\text{lead}^\text{after}=-x_\text{lag}^\text{before}$ and \textit{vice versa}.
In principle, 
a propagating displacement pulse of the same type but without damping should also be possible in 
low-Reynolds-number flow, as swapping trajectories 
exist 
in this regime as well \cite{reddig_nonlinear_2013}.

\begin{figure}%
	\centering
	\includetikz{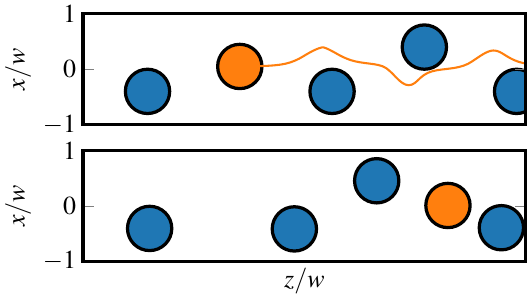}
	\caption{
		Top: Example of a passing trajectory in a staggered train when the initial 
		lateral displacement from the equilibrium position is too large.
		Bottom: Snapshot after the particle has passed two of its neighbors.
		See also video 7 in the supplemental material.
	}
	\label{fig:rigid_multi_phonon_passing}
\end{figure}

We also mention that if the displacement brings the first particle close to or across the channel centerline 
such that $\Delta |x|/w\leq -0.4$,
it becomes too fast and can no longer swap its position with the next particle. 
Instead, it moves through the staggered train (see \prettyref{fig:rigid_multi_phonon_passing}) and leaves behind a defect, where two neighboring particles move on the same streamline. 
This is reminiscent of the passing trajectory for a particle pair identified in Ref.\ \cite{schaaf_flowing_2019}.

\subsubsection{Quantitative analysis of the 
displacement pulse}

\begin{figure}%
	\centering
	\includetikz{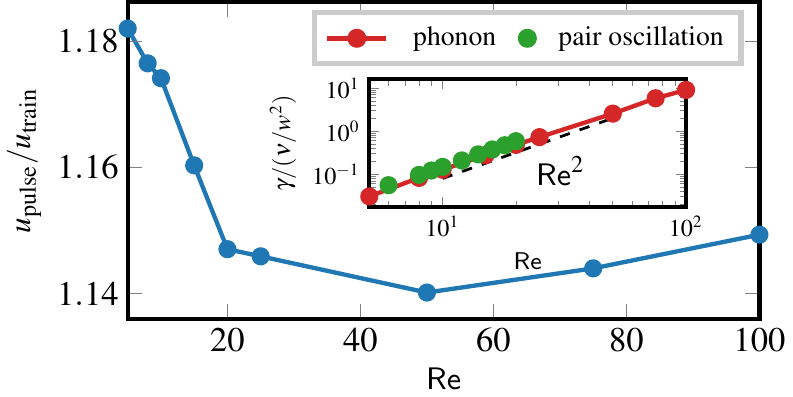}
	\caption{
		Ratio of pulse to train velocity (main plot)
		and damping rate $\gamma$ 
		(inset)
		of 
        the displacement pulse as a function of the Reynolds number. 
        In all cases the initial axial spacing 
        is $\Delta z_0/a\approx 4.2$ and the initial displacement 
        is $\Delta |x|_0 / w = -0.2$ toward the channel center. 
        For comparison, we also plot $\gamma$ for an oscillating particle pair from \cite{schaaf_flowing_2019}.
}
	\label{fig:rigid_multi_phonon_frequency_damping}
\end{figure}

In \prettyref{fig:rigid_multi_phonon_frequency_damping} we plot the ratio of pulse velocity to train velocity $u_{\mathrm{pulse}} / u_{\mathrm{train}}$ (main plot)
and the damping rate $\gamma$ (inset) of the displacement pulse as a function of the Reynolds number $\re$. 
This is similar to
our analysis of
the damped oscillations for a pair of rigid particles~\cite{schaaf_flowing_2019}.
The velocity ratio is roughly constant in $\mathrm{Re}$ so that we identify a linear dependence $u_{\mathrm{pulse}} \propto u_{\mathrm{train}} \propto \mathrm{Re}$, which makes sense since the fluid flow directly determines how fast displaced particles approach their neighbors.
A similar scaling was observed for the oscillating frequency of a pair of rigid particles~\cite{schaaf_flowing_2019}.
The pulse velocity is also larger than the train velocity since the displacement pulse is propagated by faster moving particles.
For the damping rate $\gamma$ of the 
displacement pulse in the inset we find good agreement with the damping rate of the oscillating particle pair
in the regime of $\re\leq 20$.
The $\gamma$ values for the propagating pulse are slightly lower. 
The damping rate scales quadratically with $\re$, only
for higher Reynolds numbers the scaling deviates slightly from $\re^2$.
Thus, damping of the propagating pulse is a clear inertial effect due to the acting inertial lift force and therefore the scaling with $\re^2$ is expected.

\begin{figure}%
	\centering
	\includetikz{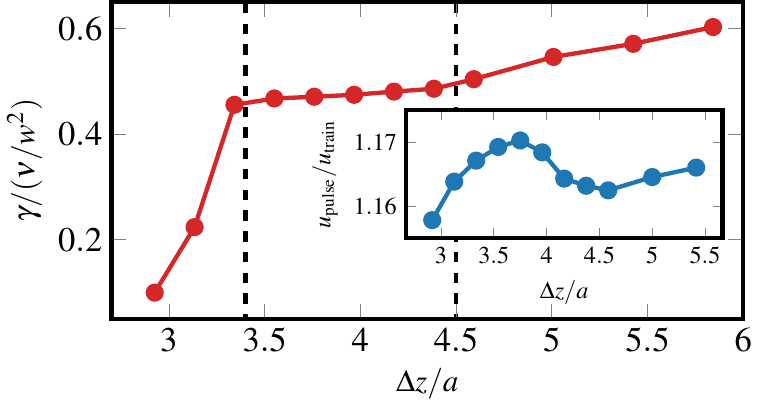}
	\caption{
		Damping rate $\gamma$ (main plot) and ratio of pulse to train velocity (inset) of the displacement pulse as a function of axial particle distance $\Delta z$.
		The other 
        parameters are $\re=20$ and $\Delta |x|_0/w=-0.2$.
        The dashed lines indicate a region with almost constant $\gamma$.
	}
	\label{fig:rigid_phonon_frequency_damping_vary_dz}
\end{figure}

In our setting using periodic boundary conditions along the channel axis, we can compress or expand the infinitely 
extended particle train
by changing the channel length. 
This allows to study pulse propagation under tension.
A possible experimental realization are expanding or contracting channels, where the channel width changes abruptly. 
For example, in an expanding channel the particle spacing relaxes slowly to its larger distance \cite{ lee_dynamic_2010,deng_inertial_2017}.
In \prettyref{fig:rigid_phonon_frequency_damping_vary_dz} we show ratio of pulse to train velocity 
$u_{\mathrm{pulse}} / u_{\mathrm{train}}$ (inset)
and the damping rate $\gamma$ (main plot) as a function of the axial particle distance $\Delta z$. 
The velocity ratio again is nearly constant with varying $\Delta z$, while the train velocity increases linearly with $\Delta z$ due to the increased friction (not shown).
For the damping rate we find three different regimes.
When the staggered train is strongly compressed,
the damping rate is strongly reduced, which is due to the strong repulsive interactions between particles.
Additionally, we observe that the displacement pulse no longer travels in one direction only, rather it propagates in both directions at the same time as we demonstrate in video 8 in the supplemental material.
For axial distances around the equilibrium value ($3.5<\Delta z/a<4.5$) the damping rate
is almost constant and
it slightly increases 
for $\Delta z/a>4.5$.
Here, the regular train is not stable. 
Instead, the distances between two particles contract
leaving larger gaps 
between particle pairs.
Since then the particles relax toward their equilibrium positions more like an individual particle, 
the overall damping rate increases.

\begin{figure}%
	\centering
	\includetikz{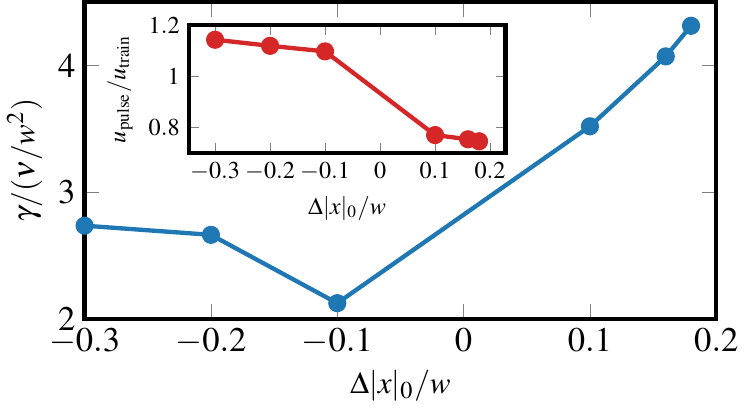}
	\caption{
		Damping rate $\gamma$ (main plot) and ratio of pulse to train velocity (inset) of the displacement pulse as a function of the initial displacement $\Delta |x|_0$. 
		The other parameters are $\re=20$ and $\Delta z/a=4.2$.
}	
	\label{fig:rigid_phonon_frequency_damping_vary_dx}
\end{figure}

Finally, we also varied the initial displacement $\Delta |x|_0$ from the equilibrium position and plot in \prettyref{fig:rigid_phonon_frequency_damping_vary_dx} ratio of pulse to train velocity $u_{\mathrm{pulse}} / u_{\mathrm{train}}$ (inset) and the damping rate $\gamma$ (main plot).
Positive $\Delta |x|_0$ means that the particle is moved toward the channel wall. 
Since it thereby slows down relative to the staggered train, it approaches the particle behind and the displacement pulse moves against the staggered train.  
For the velocity ratio in the inset this means, while for positive and negative $\Delta |x|_0$ the pulse velocity is relatively constant, the ratio $u_{\mathrm{pulse}} / u_{\mathrm{train}}$ jumps from a value larger than one to a value smaller than one when $\Delta |x|_0$ becomes positive.
The dependence of the damping rate is less intuitive. 
It is constant for large negative $\Delta |x|_0$, goes through a minimum and then increases linearly from  $\Delta |x|_0=-0.1$ for increasing $\Delta |x|_0$.

\subsubsection{Influence of a defect on the pulse propagation}

\begin{figure}[t!!]
	\centering
	\includetikz{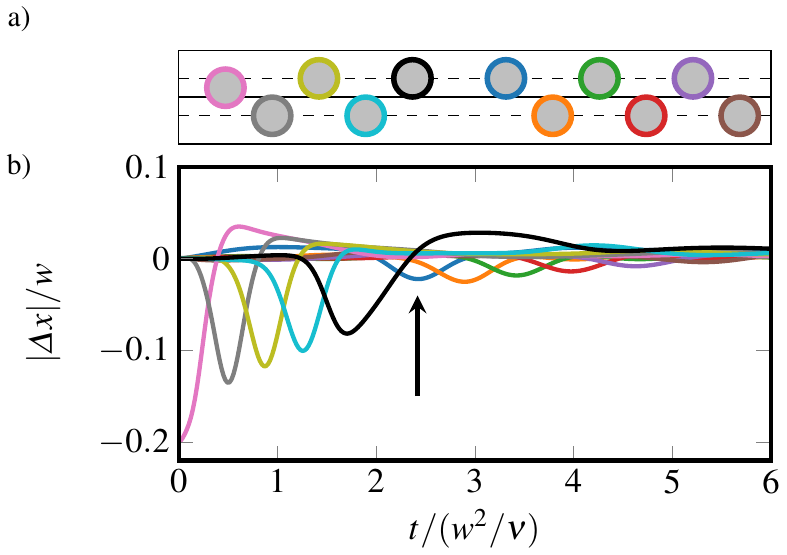}
	\caption{
		(a) Staggered particle train with a defect and an initial displacement of the first particle counted from the end (pink).
		The equilibrium axial particle distance is $\Delta z /a = 4.2$ and $\Delta z /a = 8.4$ between the defect particles.
		(b) Lateral particle displacement from the equilibrium position, $\Delta |x(t)|=|x(t)| - x_\text{eq}$,	plotted versus time for all the particles in the upper and lower channel half.
		The color coding is the same as in (a). 
		The arrow indicates the strongly damped displacement of the first particle after the defect has been passed.
	}
	\label{fig:rigid_phonon_defect}
\end{figure}

At the end we study how a defect in the staggered particle train strongly dampens the propagating pulse and show our results in \prettyref{fig:rigid_phonon_defect}. 
We
reduce the number of particles to 11 and introduce a defect with two neighboring particles on the same streamline, as already investigated above.
As illustrated in \prettyref{fig:rigid_phonon_defect}, the pulse is initiated
{at the pink particle at the left end of the channel, which is}
the fifth particle to the left of the defect.
Up to the defect {(black particle) the pulse}
propagates
as before. 
However, having passed the defect {(blue particle)} it is strongly damped
such that the pulse vanishes almost completely.

\section{Conclusions}
The axial alignment of particles in an inertial microchannel is an important feature for the counting, manipulation, and sorting of cells. 
Therefore, the stability of trains is a crucial ingredient for designing and understanding lab-on-a-chip devices. 
As the literature on this topic does not always agree with their findings, we focused on a detailed analysis for one specific set of parameters.

The stability of cross-streamline pairs is accepted in the literature. 
We show that particles in such a pair 
attract each other over large distances while 
their lateral positions 
hardly change. 
A cross-streamline pair always contracts or expands to its equilibrium axial distance.
For same-streamline pairs we thoroughly analyze and thereby confirm the 
result of our previous work that the same-streamline configuration 
does not have
an equilibrium axial spacing~\cite{schaaf_flowing_2019}. 
However, from smaller distances it quickly expands to a characteristic separation but even at long times very slowly drifts apart.
Their dynamics is dominated by a repulsive interaction, which
rapidly decays with increasing axial distance.
With our simulations we also reproduce the distribution of axial distances for a same-streamline pair measured in experiments \cite{lee_dynamic_2010}. In particular, it has a well-defined peak at about twice the distance of cross-streamline pairs.

Then, we extended our analysis to particle trains with more than two particles and first analyzed how staggered trains relax to their equilibrium configurations. 
In particular,
a staggered train initialized with an axial particle spacing larger than the equilibrium distance contracts non-uniformly.
The non-uniform contraction
is related to two effects of 
collective drag reduction: (i) when two particles in a cross-streamline configuration approach each other they slow down since they exhibit less resistance to the driving Poiseuille flow and (ii) the center-of-mass velocity of staggered trains decreases the more particles are in the train (\textit{peloton effect}).
These two effects drive the non-uniform contraction
in the front and 
back of a staggered train. 
While in the front the leading pair slows down and collects more and more particles, in the back trailing pairs of particles separate from the rest of the train. 
Only 
with time the pairs catch up with the particle train in front of them and form one large train.
Finally, we find a master curve for the axial spacing within a staggered train. 
The spacing between the particles increases from the back to the front of the train and ultimately saturates for sufficiently long trains.
So a staggered train is slightly expanded at the front relative to the back.
In experiments a particle train slowly %
expands when it enters a channel with a suddenly expanding cross section \cite{lee_dynamic_2010,deng_inertial_2017}.
This could be a means to observe the scenario outlined here.

For linear trains we find a very similar behavior as for same-streamline pairs.
Starting from a particle distance similar to the staggered train, the
spacing relaxes in the beginning phase to a value close to experimental results \cite{kahkeshani_preferred_2016} about twice the distance of cross-streamline pairs. 
Then, the particles  continue to slowly drift apart non-uniformly.
The leading particle moves the fastest and separates from the rest of the train.
This confirms a similar observation reported by Gupta \textit{et al.} for a smaller confinement ratio~\cite{gupta_conditional_2018}.
In addition, we are able to reproduce the statistics of particle distances observed in experiments \cite{kahkeshani_preferred_2016}.

Finally, we 
analyzed how a displacement pulse migrates as inertial microfluidic phonon through a staggered {train}. 
The displacement is transported from one particle to another via swapping trajectories, where the inertial lift forces 
damps the amplitude of the displacement pulse.
When the initial 
displacement is too large, the displaced particle itself moves through the staggered train and leaves behind a defect.
The ratio of pulse to train velocities is almost constant for varying Reynolds number and axial particle distance, whereas it is by ca. 30\% smaller for initial displacement toward the wall compared to perturbations toward the channel center.
For the damping rate of the displacement pulse we confirm the quadratic scaling with the Reynolds number, which identifies
damping as an inertial effect.
The damping rate is almost constant for
varying 
axial distance between the particles or changing
line density.
Only when the particles are close together the damping rate is reduced. 
This is especially 
interesting when the line tension
of staggered trains changes rapidly
when entering sections of the channel with expanding or contracting  cross sections 
\cite{lee_dynamic_2010,deng_inertial_2017} since then the trains want to expand or contract.

We hope that with our careful numerical study we initiate further research on staggered and linear particle trains to clarify some of the still open questions and also to provide guidance how microfluidic phonons in the inertial regime behave.

\begin{acknowledgements}
We appreciate helpful discussions with Timm Krüger.
We acknowledge support from the Deutsche Forschungsgemeinschaft in the framework of the Collaborative Research Center SFB 910.
\end{acknowledgements}

\bibliographystyle{spphys}
\bibliography{literature}

\end{document}